\documentstyle[emulateapj]{article}
\def\Meszaros{M\'esz\'aros~}

\begin{document}

\title{GRB Precursors in  the Fallback Collapsar Scenario
 }

\author{Xiang-Yu  Wang\altaffilmark{1,2},   Peter
M\'esz\'aros\altaffilmark{1,3}}

\altaffiltext{1}{Department of Astronomy and Astrophysics,
Pennsylvania State University, University Park, PA 16802, USA}
\altaffiltext{2}{Department of Astronomy, Nanjing University,
Nanjing 210093, China} \altaffiltext{3}{Department of Physics,
Pennsylvania State University, University Park, PA 16802, USA}

\begin{abstract}
Precursor emission has been observed in a non-negligible fraction
of gamma-ray bursts.The time gap between the precursor and the
main burst extends in some case up to hundreds of seconds, such as
in GRB041219A, GRB050820A and GRB060124. Both the origin of the
precursor and the large value of the time gap are controversial.
Here we investigate the maximum possible time gaps arising from
the jet propagation inside the progenitor star, in models which
assume that the precursor is produced by the jet bow shock or the
cocoon breaking out of the progenitor. Due to the pressure drop
ahead of the jet head after it reaches the stellar surface, a
rarefaction wave propagates back into the jet at the sound speed,
which re-accelerates the jet to a relativistic velocity and
therefore limits the gap period to within about ten seconds. This
scenario therefore cannot explain gaps which are hundreds of
seconds long. Instead, we ascribe such long time gaps to the
behavior of the central engine, and suggest a fallback collapsar
scenario for these bursts. In this scenario, the precursor is
produced by a weak jet formed  during the initial core collapse,
possibly related to MHD processes associated with a short-lived
proto-neutron star, while the main burst is produced by a stronger
jet fed by fallback  accretion onto the black hole resulting from
the collapse of the neutron star. We have examined the propagation
times of the weak precursor jet through the stellar progenitor. We
find that the initial weak jet can break out of the progenitor in
a time less than ten seconds (a typical precursor duration)
provided that it has a moderately high relativistic Lorentz factor
$\Gamma\ga10$. The longer ($\sim 100$ s) time gap following this
is  determined, in this scenario, by the fall-back timescale,
which is at the same time long enough for the exit channel to
close after the precursor activity ceases, thus allowing for the
collimation by the cocoon pressure of the subsequent main jet, as
required.

\end{abstract}

\section{Introduction}

Weak, softer precursor emission, occurring from several seconds to
hundreds of seconds before the main gamma-ray pulse, is
occasionally detected in long gamma-ray bursts (GRB).  Often there
is almost no emission detectable in the time gap between the
precursor and the main event. The earliest detections of precursor
events were made with the Ginga satellite (GRB900126, Murakami et
al. 1991). Since then, precursors have been found in many bursts,
e.g. by BeppoSAX (e.g. GRB011121, Piro et al. 2005), HETE2 (e.g.
GRB030329, Vanderspek et al. 2004), INTEGRAL (e.g.  GRB041219A,
Mereghetti et al. 2004) and Swift (e.g. GRB050820A, Cenko et al.
2006; GRB060124, Romano et al. 2006; GRB061121, Page et al. 2007).
Lazzati (2005) analyzed a sample of bright, long BATSE burst light
curves in the 200 s period before the detection of the GRB prompt
emission, and found that a sizable fraction ($\sim20\%$) of them
show evidence for gamma-ray emission above the background between
10 to $\sim200$ s (the limit of the investigated period) before
the main burst. The precursors are typically characterized by
non-thermal spectra which tend to be softer than the main burst,
and contain from $0.1 \%$ to a few $\%$ of the total event counts.
For some bursts with known redshifts, the isotropic energies of
the precursors are determined, and can be as large as $E_{\rm
iso}\simeq5\times10^{51}{\rm erg}$ for GRB060124 (Romano et al.
2006; Misra et al. 2007). The time gap between the precursor and
the main event ranges from  several seconds to a few hundreds of
seconds (e.g. 200 s for GRB041219A and GRB050820A, and 570 s for
GRB060124 in the observer frame).

Theoretical models for precursors have been proposed, relating it
to the breakout of the main GRB jet. As suggested by Ramirez-Ruiz
et al. (2002a) and Waxman \& \Meszaros (2003), when the jet is
making its way out of the stellar mantle, a bow shock runs ahead
and a strong thermal precursor is produced as this shock breaks
out{\footnote{A non-thermal component could also arise during
shock breakout, due to the repeated bulk Compton scattering of
shock breakout thermal photons by the relativistic ejecta driving
the shock (Wang et al. 2006). } }. In this model, as usual, the
main gamma-ray burst is produced when the internal shock occurs at
some radius $R_\gamma$, the delay time between precursor and the
main burst being $\Delta{t}\simeq R_\gamma/2\Gamma^2 c$, where
$\Gamma$ is the Lorentz factor of the jet. Since the bursts with a
precursor also display significant $\ga{\rm MeV}$ emission,
similarly to  other typical long bursts, their bulk Lorentz
factors must be likewise $\Gamma \ga 100$ (e.g. Baring \& Harding
1997; Lithwick \& Sari 2001), hence the typical delay $\Delta t$
is less then a second for the typical internal shock radii
$R_\gamma\sim10^{12}-10^{14}{\rm cm}$ (see \Meszaros (2006) and
Zhang (2007) for a recent review). Since the variability timescale
is also $R_\gamma/2\Gamma^2 c$, this scenario would imply a gap
timescale comparable to the variability timescale, which is
clearly not the case. Morsony et al. (2007) investigated the jet
propagation through the star with an elaborate numerical
simulation and identified three distinct phases during the
evolution, i.e. the precursor phase, the shocked jet phase and the
unshocked jet phase. They find that a high-pressure cocoon, formed
as the sub-relativistic jet head makes its way out of the star, is
released  when the head of the jet reaches the stellar surface
(which can produce a precursor, Ramirez-Ruiz et al. 2002b). Within
the framework of this model, an observer which is located at small
or modest viewing angles relative to the jet axis would be
expected to see first a relatively bright precursor, then a gap
phase during which there is little emission, and finally a bright
GRB phase when the unshocked jet reaches the radius of the stellar
surface.

In \S 2, we consider first a precursor model such as discussed
above, which is based on the time delays associated with the same
jet giving rise both to the precursor and to the main burst event.
We show that the shocked jet phase could not last much longer than
$\sim 10$ s, due to the fact that once the jet head reaches the
stellar surface, a rarefaction wave will form which propagates
back into the shocked jet component, decreasing the pressure in
the shocked jet significantly, and as a result, the jet attains
its ultimate relativistic velocity. Hence, the propagation time
taken by the unshocked jet to reach the stellar surface will be
limited to $\sim R_{\star}/c$, which is only from a few seconds to
ten seconds for Wolf-Rayet progenitors of long GRBs, where
$R_{\star}$ is the stellar radius of the progenitor. Thus it
appears that this model has difficulties in explaining the
$\sim100$ s long gaps seen in some GRBs. In \S 3 we then suggest
that gaps hundreds of seconds long could arise in a fallback
collapsar scenario, or in a ``type II collapsar'' scenario
(MacFadyen et al. 2001), { the gap time being related to the
feeding time of the fallback disk, or the instability timescale in
the accretion disk. We argue that in such scenarios an initial
weak jet, which is still able to break} out through the star,
produces the precursor, while the main burst is produced by a
subsequent more energetic jet, fed by accretion of the fallback
gas. { The initial precursor jet, whatever its formation
mechanism,  must be weak enough not to disrupt the star, so that a
fallback disk can still form later on. This constrains the
luminosity of the precursor jet to be lower than $E_{\rm b}/t_j$,
where $E_{\rm b}\simeq 10^{51}{\rm erg}$ is the binding energy of
the star envelope and $t_j$ is the precursor duration. Motivated
by the findings of Woosley \& Zhang (2007) that a weaker jet
propagates more slowly inside the jet,  in \S 4 we study
analytically the dependence of the propagation time of the jet on
other jet parameters. We confirm here the findings of these
simulations, and  obtain some constraints on the precursor jet
luminosity and Lorentz factors. } Some possible formation
mechanisms for the precursor jet are considered in \S
\ref{sec:jetmech}. A summary of the results is given in \S
\ref{sec:discussion}.

\section{Jet dynamics inside the star and at the breakout}
\label{sec:jetdyn}

In the collapsar model (Woosley 1993; Paczy\'{n}ski 1998;
MacFadyen \& Woosley 1999) , GRBs are caused by relativistic jets
expelled along the rotation axis of a collapsing stellar core. The
relativistic jets are due to a black hole or a neutron star
accretion disk, after the iron core of the massive star progenitor
has collapsed.

As the jet advances through the star, it drives a bow shock ahead
of itself into the star, while the ram pressure of the shocked gas
ahead of the jet drives a reverse shock into the head of the jet,
slowing the jet head down to a sub-relativistic velocity. Thus
there are three distinct regions: 1) In front of the contact
discontinuity between the jet and the stellar gas, there is a thin
layer of shocked stellar gas moving ahead with a sub-relativistic
velocity $\sim v_h$ into the star; 2) behind the contact
discontinuity, there is a shocked jet region, where the
relativistic jet with $\Gamma_j\gg1$, is slowed down to a velocity
$\sim v_h$ by the reverse shock, and 3) and below this is the
unshocked jet, whose bulk Lorentz factor behaves as if it were a
free jet. In this unshocked jet,  at lower radii, the gradual
conversion of its internal energy into kinetic energy results in
$\Gamma_j=\Gamma_0 (r\theta_{j}/r_0\theta_{0})$, until a
saturation radius at which it reaches its asymptotic value. Here
$\theta_j$ is the opening angle of the jet at radius $r$ and
$\Gamma_0$ and $\theta_0$ are, respectively, the initial Lorentz
factor and opening angle at the injection radius $r_0$.

Let us first consider the scenario in which the precursor is
related to the jet breakout (e.g. Ramirez-Ruiz et al. 2002a,b;
Waxman \& \Meszaros 2003; Morsony et al. 2007). In this scenario,
the precursor is produced by the bow shock emission or the cocoon
emission when the jet head breaks out of the  stellar surface.
After this, the shocked jet phase emerges, which may have a
relatively small opening angle for ultra-relativistic material due
to that  the mixing of the shocked jet with stellar material
lowers its terminal Lorentz factor, so that an observer located at
a  larger viewing angle would see a dark phase. The relativistic
reverse shock may be located well below the stellar surface at the
time when the jet head has just reached the surface. One may think
that this dark phase could last long enough, if the shocked jet
material moves with a low velocity. However, once the jet head
reaches the stellar surface, the pressure in front of the jet head
decreases suddenly and a rarefaction wave will form and propagate
back into the shocked jet material at the speed of sound. The
speed of sound in the shocked jet plasma is relativistic,
$c_s=c/\sqrt{3}$, even if the head velocity $v_h\ll c$. When this
rarefaction wave arrives at the reverse shock, the pressure of the
shocked jet material also drops and it can no longer decelerate
the fast unshocked jet. Supposing that the reverse shock is
located at a position well below the stellar surface when the jet
head reaches the stellar surface, the width of the shocked jet is
$\Delta\la R_{\star}$, and the time that the rarefaction wave
takes to arrive at the reverse shock is $t_1=\Delta/c_s\la
R_{\star}/c_s=6R_{\star,11} \,{\rm s} $. A new system of forward
and reverse shocks will form, and the rarefied shocked jet plasma
is accelerated to a relativistic velocity with a Lorentz factor
\begin{equation}
\Gamma_{h2}=\Gamma_{h1}[4\Gamma_{h1}(1+c_s t/\Delta)^3]^{1/4}
\end{equation}
where $\Gamma_{h1}$ is the Lorentz factor of the shocked jet
before the rarefaction wave propagating back (see Eq.(15) of
Waxman \& \Meszaros 2003). Even for a sub-relativistic velocity
$\Gamma_{h1}\simeq1$, the rarefied shocked jet is accelerated to a
relativistic velocity with $\Gamma_{h2}\ga2$ after a time $t_1$.
The unshocked jet then takes a time $t_2\simeq
R_{\star}/c=3R_{\star,11} \,{\rm s}$ to reach the stellar surface.
Supposing that the internal shock occurs at $R_\gamma$ where it
produces the burst emission, the additional delay relative to the
precursor emission is $t_3=R_\gamma/2\Gamma^2 c$. Therefore the
total gap period between the precursor and the main burst would be
only
\begin{equation}
T_{\rm gap}=t_1+t_2+t_3\simeq 10 R_{\star,11}\, {\rm s}.
\end{equation}

\section{Fallback collapsar scenario }
\label{sec:fallback}

The above described difficulty in interpreting long ($\sim 100$ s)
precursor gaps through a single jet breakout makes it necessary to
explore other alternatives. Here we propose a model where the gap
is related to the central engine activity, and the precursor is
produced by an initial weak  jet launched before the main jet. The
$\sim100 {\rm s}$ long gap is remiscent of the natural timescales
calculated numerically in the fallback collapsar (type II
collapsar) model (MacFadyen et al.  2001), where the fallback disk
forms minutes to hours after the initial core
collapse{\footnote{Cheng \& Dai (2001) suggested a similar,
``two-step'' jet model, which attempts to solve the baryon
contamination problem in the hypernova model of GRBs.}}. In this
scenario, the collapse of the iron core initially forms a
proto-neutron star and launches a supernova shock. However, this
shock lacks sufficient energy to eject all the matter outside the
neutron star, especially for more massive helium cores, because
the gravitational binding energy of the helium core increases with
mass roughly quadratically, while the explosion energy decreases
(Fryer 1999). If the explosion energy is lower than $10^{51}{\rm
erg}$, as the supernova shock decelerates while it travels
outward, some of the expanding material can decelerate below the
escape velocity and falls back.

During the initial collapse, it is possible that a weak jet with
an energy of, say, a few times $10^{50}{\rm erg}$, is produced as a
result of MHD processes in the collapsed core, or through propeller
effects associated with a proto-neutron star, etc. (e.g. Wheeler et
al, 2000). If such a jet is weak enough so as not to immediately
disrupt the star, a fallback accretion disk can form in the core on
a fallback timescale, which can launch a stronger jet producing the
main burst.  We argue that such initial weak jets may be responsible
for precursors with non-thermal spectra, through internal dissipation
mechanisms such as internal shocks or reconnection, after they exit
the stellar progenitor. The non-thermal spectra are consistent with
those observed from most GRB precursors (e.g.  Lazzati 2005;  Cenko et
al. 2006; Romano et al. 2006; Page et al.  2007).

After a time $\sim 100$ s following the initial collapse, in the
case of a weak supernova, enough gas has fallen back for the
neutron star to collapse to a black hole. Gas that continues to
fall back with sufficient angular momentum will settle into a disk
(MacFadyen et al. 2001).  The black-hole accretion disk system may
then produce stronger jets which, through neutrino annihilation or
MHD processes (e.g. Blandford-Znajek mechanism), can power the
main burst event.

\section{Jet propagation through the stellar progenitor}
\label{sec:jetprop}

We investigate now whether a weak precursor jet of luminosity
$L_j\sim10^{49}{\rm erg s^{-1}}$ lasting for $\sim10$ seconds (the
precursor duration) can break out of the star.  Previous analytic
studies on the jet dynamic inside the star and the jet breakout
time have been done by, e.g. \Meszaros \& Rees 2001; Ramirez-Ruiz
et al. (2002b); Matzner (2003), Lazzati \& Begelman (2005) and
Toma et al. (2006). As recently found by Woosley \& Zhang (2007)
in the simulation of jet propagation through the star, a
lower-energy jet takes a longer time to break out of the star. A
 jet with a lower Lorentz factor may also take a longer time, while at present
there is no constraint yet on the Lorentz factor of the precursors
jets since their spectra  do not display significant $\ga$ MeV
emission. Since the precursor lifetime is usually only about ten
seconds and the breakout time must be shorter than this (otherwise
the jet will be chocked), it is not obvious whether it can really
break out of the star or not. We will show below that, for a
low-energy jet to be able to break out of the star in a time
shorter than the precursor duration, it must have a moderately
high asymptotic Lorentz factor, $\eta= L/\dot{M}c^2\ga 10$. The
asymptotic Lorentz factor of this weak precursor jet can, however,
be lower than that of the typical main GRB ejecta, which are
characterized by $\eta>100$, and this could account for the softer
spectrum of the precursor compared to the main burst.
Phenomenological relations between the energy (or luminosity) and
the spectrum of the burst have been found by e.g. Amati et al.
(2002) and Yonetoku et al. (2004), in which a lower energy (or
luminosity) burst tends to have a lower peak energy, hence a
softer spectrum.

\subsection{Jet crossing time}

The jet head velocity is given by the longitudinal balance between
the jet thrust and the ram pressure of material ahead of it
(\Meszaros \& Rees 2001; Matzner 2003)
\begin{equation}
v_h=\left(\frac{L_j}{2\pi \theta_j^2 r^2 c
\rho}\right)^{1/2}=7\times10^9
r_{11}^{1/2}L_{j,49}^{1/2}\theta_{j,-2}^{-1}\, {\rm cm s^{-1}}
\end{equation}
where $\theta_j$ is the opening angle of the jet while propagating
inside the star, whose value will be derived below, and $\rho$ is
the stellar density at radius $r$. The presupernova density
profile is assumed to be roughly described by $\rho\propto r^{-3}$
out to the edge of the He core at $r=10^{11}{\rm cm}$ (MacFadyen
et al. 2001), at which point we take the density to be $\rho=1\,
{\rm g cm^{-3}}$.  The total time taken by the jet head to move
from the interior of the star to the surface is
\begin{equation}
t=\int\frac{dr}{v_h}=\frac{\alpha r}{v_h}=14 \alpha
r_{11}^{1/2}L_{j,49}^{-1/2}\theta_{j,-2} \,\,{\rm s}
\end{equation}
where $\alpha>1$ is an integrating factor over $r$ (discussed
after Eq.(10)). While the jet is propagating inside the star with
a sub-relativistic velocity, a significant fraction of jet
``waste'' energy is pumped into the cocoon surrounding the
advancing jet. The unshocked jet moving with $\Gamma_j\gg1$ is now
constrained in the transverse direction by the pressure of the
cocoon, so that it gets collimated and becomes more penetrating.
The opening angle $\theta$ of the unshocked jet is determined by
the balance between the pressure of the unshocked jet in the
transverse direction and that of the cocoon.

The energy stored in the cocoon is roughly $E_c\simeq L_j (t-r/c)$
when the sub-relativistic jet reaches radius $r$  after a time
$t$. The volume of the cocoon is $V_c=\frac{\pi}{3}r_{\bot}^2r$,
where $r_{\bot}=v_{\bot}t$ is the transverse size of the cocoon
(Matzner 2003). Since the cocoon pressure is much larger than that
of the stellar material, the transverse expansion velocity
$v_{\bot}$ of the cocoon is given by the balance between the
cocoon pressure $p_c$ and the ram pressure of the stellar
material, i.e. $v_{\bot}=(p_c/\rho)^{1/2}$. From $p_c=E_c/3V_c$,
we  get
\begin{equation}
p_c=\left(\frac{L_j \rho}{\pi r t}\right)^{1/2}=\left(\frac{L_j
\rho v_h}{\alpha \pi r^2 }\right)^{1/2}.
\end{equation}

The pressure in the unshocked relativistic jet in the transverse
direction, from the relativistic Bernoulli equation, is
$p_{j,\bot}=\frac{1}{3}L_j/(2\pi r^2 \theta_j^2 c\Gamma_j^2)$. As
the internal energy gets converted into the kinetic energy of the
jet, $\Gamma_j$ increases. Depending on whether the jet Lorentz
factor has  saturated or not before the jet breaks out of the
star, there are two cases for the evolution of $\Gamma_j$: one is
the high $\eta$ case (case I), in which $\Gamma_j\propto
(r\theta_j)=\Gamma_0 (r/r_0)(\theta_j/\theta_0)<\eta$ all along
the path (see the left panel of Fig. 1) , and the other is the low
$\eta$ case (case II) in which $\Gamma_j$ has reached its
saturation value $\eta$ at some radius inside the star (see the
right panel  of Fig. 1). The latter case is more easily satisfied
when $\eta$ is lower or  when the initial injection radius $r_0$
is smaller. The jet pressure is different in the two cases, being
given by, respectively,
\begin{equation}
 p_{j,\bot}=  \cases{
 {L_j r_0^2 \theta_0^2}/(6\pi c r^4 \theta_j^4 \Gamma_0^2)  & $\Gamma_j<\eta$ (\rm Case I); \cr
 {L_j}/(6\pi c r^2 \theta_j^2 \eta^2) & $\Gamma_j=\eta$ (\rm Case II) \cr}
\label{eq:pjet}
\end{equation}
From $p_{j,\bot}=p_c$, we get the opening angle for the two cases
respectively,
\begin{equation}
\theta_j= \cases{
 (\alpha^2 L_j/648\pi \rho
c^3)^{1/14}(r_0\theta_0/\Gamma_0)^{4/7} r^{-5/7}; \cr
(\alpha^2L_j/648\pi r^2 \rho c^3)^{1/6}\eta^{-4/3} \cr}
\end{equation}
For case I, with  the parameters $\theta_0=1$, $\Gamma_0=1$,
$r_0=10^7 {\rm cm}$ and the stellar density profile $\rho=1
(r/10^{11}{\rm cm})^{-3} \,{\rm g cm^{-3}}$, the opening angle is
\begin{equation}
\theta_j=1.5\times10^{-3}\alpha^{1/7}
L_{j,49}^{1/14}r_{11}^{-1/2}r_{0,7}^{4/7},
\end{equation}
while for  case II,
\begin{equation}
\theta_j=3\times10^{-3}\alpha^{1/3}L_{j,49}^{1/6}r_{11}^{1/6}\eta_1^{-4/3}.
\end{equation}
Note that these values of the opening angles are the ones at the
jet breakout time, which are not the final values that we expect
in the internal shock phase or in the afterglow. After exiting the
star, the hot jet is expected to expand freely to an opening angle
comparable to $1/\Gamma_{\rm br}\sim 0.1$, where $\Gamma_{\rm br}$
is the Lorentz factor of the jet at the breakout time.

With the above values of the opening angles, we can obtain the jet
breakout time from  Eq.(4) for the two cases,
\begin{equation}
 t_{\rm br}=  \cases{
 \alpha^{8/7}r^{9/7}(\frac{4\pi^3\rho^3 c^2}{L_j^3})^{1/7}(\frac{r_0 \theta_0}{\Gamma_0})^{4/7}
 =2\alpha^{8/7}L_{j,49}^{-3/7}r_{0,7}^{4/7}{\rm s}; \cr
 \alpha^{4/3}r^{5/3}(\frac{L_j}{\pi \rho})^{-1/3}\eta^{-4/3}=4\alpha^{4/3}r_{11}^{2/3}L_{j,49}^{-1/3}\eta_1^{-4/3} {\rm s} \cr}
\label{eq:tbreak}
\end{equation}
Here  $\alpha$ is the integrating factor over $r$ in Eq.(4). In
case I, since $\rho\propto r^{-3}$ starts roughly above
$r=10^9{\rm cm}$, below which the free fall density profile
$\rho\propto r^{-3/2}$ may apply (\Meszaros \& Rees 2001),
$\alpha=_1{\rm ln}(r/10^9{\rm cm})\simeq 4$ for $r=10^{11}{\rm
cm}$. In case II, the integration over $r$ in Eq.(4) gives
$\alpha_2=3/2$. We can see that a low-luminosity, weak jet takes a
longer time to break out of the star, consistent with recent
findings by Woosley \& Zhang (2007) from the numerical
hydrodynamic simulation of the jet propagation. The dependence of
the breakout time $t_{\rm br}$ on the luminosity is illustrated in
the inset plot of  the left panel of Fig.1. Requiring that the
breakout time $t_{\rm br}$ is less than the duration of the
precursor $T_p\simeq 10 \,{\rm s}$, we get, for case I,
\begin{equation}
 L_j\ga10^{49}(\alpha_1/4)^{8/3}r_{0,7}^{4/3}T_{p,1}^{-7/3} {\rm erg
 s^{-1}}.
 \end{equation}
The asymptotic Lorentz factor $\eta$ in this case is required to
be  larger than $\Gamma_{\rm br}=\Gamma_0(r\theta_j/r_0\theta_0)
\simeq 15$, from the high $\eta$ assumption of case I. For case
II, the breakout time is  sensitive to $\eta$, more than to the
jet luminosity, and we get
\begin{equation}
\eta\ga10(\alpha_2/1.5)r_{11}^{1/2}L_{j,49}^{-1/4}T_{p,1}^{-3/4}
\end{equation}
by requiring the breakout time  to be shorter than $T_p\simeq10
{\rm s}$ (see  the inset plot of the right panel of Fig.1).
Therefore we  see that in both cases, it is only when $\eta\ga10$
that the jet breakout time can be $\la 10$ seconds, for a
luminosity of the order of $L_j\sim10^{49}{\rm erg s^{-1}}$. This
reflects the fact that a lower Lorentz factor jet is less
penetrating and needs a longer time to break out of the stellar
progenitor. This shows that even the precursor jet should have a
relatively high Lorentz factor, although it need not be as large
as $\ga100$ as in the case of the main burst.

\subsection{Can the precursor channel close up again?}

After the weak jet breaks out, the cocoon material escapes. There
is now a wide funnel opened by the weak jet and its accompanying
cocoon along its path. As evidenced by GRB050820A (Cenko et al.
2006), a burst with precursor, the main burst jet is collimated to
$\theta\sim0.1$, like other typical long bursts without
precursors. Since a closed channel is a necessary condition for
the main jet to be collimated, one may ask whether the initial
channel, made by the precursor jet, will have had time to close
again by the time that the main jet starts to make its way out,
$\sim100$ seconds later.

As the weak jet propagates outward, the cocoon pushes the
stellar material sideways. The opening angle of the cocoon, which
is also the funnel opening angle after the cocoon flows out, is
approximately given by
\begin{equation}
\theta_f=\theta_c=\frac{r_{\bot}}{r}=\left(\frac{L_j t^3}{\pi \rho
r^5}\right)^{1/4}=\alpha \Gamma_{\rm br}^{-1},
\end{equation}
where the last equality is obtained with the help of the second
line of Eq.(10). For $\Gamma_{\rm br}\simeq10$, the cocoon opening
angle is about tens of degrees.

The cocoon can drive a transverse shock into the stellar material
along its passage, heating it up, so the pressure of the shocked
stellar material ($p_e\sim \rho v_{\bot}^2$) is comparable to that
of the cocoon, i.e. $p_e\simeq p_c$. After the cocoon flows out,
this external stellar material will re-expand back into the funnel
at a speed of sound given by
\begin{equation}
c_s=\sqrt{p_e/\rho}\simeq\sqrt{p_c/\rho}=v_{\bot}.
\end{equation}
This means that in a time comparable to the precursor jet breakout
time, the channel closes up again. From Fig.5 of Lazzati (2005), it is
known that the precursor duration is always shorter than the gap
period of time. Therefore, we conclude that the channel opened by
the precursor jet is indeed closed again before the main jet
starts to propagate out.

The free-fall time of the material along the rotation axis is
\begin{equation}
t_{\rm ff}=3000 \left(\frac{\rho}{ 1 \,{\rm g
cm^{-3}}}\right)^{-1/2} {\rm s}.
\end{equation}
For the C/O core material of the collapsed progenitor, whose
density is $\rho\sim 500 {\rm g cm^{-3}}$,  its free-fall time is
comparable to the $\sim 100 \,{\rm s}$ gap period and therefore
this material will be essentially evacuated. However, for He core
material with $\rho\sim 1~{\rm g cm^{-3}}$, the free-fall time is
much longer than the gap period, and the funnel in this { part
(aside from reclosing) remains almost the same as before the
collapse, by the time that the main jet starts to propagate out.}

\subsection{Transient thermal emission from the open funnel}

The channel made by the initial precursor jet can, while it
remains open, produce an interesting emission signature. As the
cocoon expands transversally into the stellar material, it will
drive a shock.  Since this shock takes place in the highly
optically thick interior of the progenitor star,  it will become a
radiation-dominated shock, similarly to the case of the supernova
shock propagating inside the star (Weaver 1976).   The velocity of
the shock driven by the cocoon into the He core material can be
derived from $v_s=(p_c/\rho)^{1/2}=v_\perp$, where $p_c$ is given
by Eq.(5). This gives a shock velocity $v_s\simeq10^9{\rm cm
s^{-1}}L_{j,49}^{3/8}\theta_{j,-1}^{-1/4}r_{11}^{-3/4}\rho^{-3/8}\alpha^{-1/4}$.
For a radiation-dominated shock, this corresponds to a radiation
temperature of
\begin{equation}
T_r\sim \left(\frac{L_j\rho v_h}{a^2 \alpha\pi
r^2}\right)^{1/8}\sim 8
\alpha^{-1/8}L_{j,49}^{1/8}\left(\frac{v_h}{c}\right)^{1/8}
r_{11}^{-1/4} {\rm KeV}
\end{equation}
where $a$ is the Boltzmann  energy density constant of radiation.

To check the self-consistency of having assumed a
radiation-dominated shock, we need to compare the the radiation
diffusion length $\Delta_\gamma=c/(3n_e v_s \sigma_{\rm T})$
(where $n_e$ is the electron number density of the plasma) with
the stopping distance of the ions due to Coulomb collisios $l$.
According to Weaver \& Chapline (1974), the Coulomb friction is
the dominant dissipation mechanism controlling the shock width
(i.e. the shock is collisional), provided that the stopping
distance of the ions $l$ is larger than the radiation diffusion
length $\Delta_\gamma$. Otherwise, the shock is
radiation-dominated. For an electron temperature $T_e=T_r$, the
ion-electron collision rate is $\nu_{ie}=2\times10^{10} (\rho/1
{\rm g cm^{-3}}) (T_e/{\rm 10KeV})^{-3/2} {\rm s^{-1}}$, while the
ion-ion collision rate, for an ion temperature of $T_i=(3/16) m_p
v_s^2=200{\rm KeV}$, is $\nu_{ii}=1\times10^{10} (\rho/1 {\rm g
cm^{-3}}) (T_i/200{\rm KeV})^{-3/2} {\rm s^{-1}}$ (Spitzer 1962;
Waxman \& Loeb 2001).  Thus, the stopping  of the ions is
dominated by interaction with the electrons, with a mean free path
given by $l_{ie}=\nu_{ie}^{-1}v_s=0.05{\rm cm} (\rho/1 {\rm g
cm^{-3}})^{-1} (T_e/{\rm 10KeV})^{3/2} (v_s/10^9{\rm cm s^{-1}})$.
The radiation diffusion length is $\Delta_{\gamma}=20{\rm cm}
(\rho/1 {\rm g cm^{-3}})^{-1} (v_s/10^9{\rm cm s^{-1}})^{-1}$.
Since $\Delta_{\gamma}\gg l_{ie}$, the scale (width) of the shock
is determined by $\Delta_{\gamma}$. Except for a thin layer at the
leading edge of the shock front, the Coulomb friction is
unimportant, and the shock is effectively radiation-dominated. For
a radiation temperature of $T_r\simeq 10{\rm KeV}$, Compton
scattering gives a cooling rate of $t_{\rm
Comp}^{-1}=(8\sigma_T/3m_e c) a T_r^4= 8\times10^{10} (T_r/10{\rm
KeV})^4 {\rm s^{-1}}$, while the Bremsstrahlung cooling rate is
$t_{Brem}^{-1}=3\times10^8 (\rho/1 {\rm g cm^{-3}}) (T_e/{\rm
10KeV})^{-1/2} {\rm s^{-1}}$. Thus, Compton scattering also
provides the dominant mechanism for cooling the electrons, and
this can reduce the ion temperature significantly from those that
would occur in a Coulomb viscosity-dominated shock. These photons
diffuse ahead, creating sufficient pressure to decelerate the
electrons (as viewed in the frame of the shock front) and the ions
are then decelerated by these electrons through Coulomb friction.
We note that the above rough quantitative argument is only an
approximate treatment of this issue. A more detailed numerical
treatment taking into account the energy and momentum balance in
the shock transition as well as relativistic effects{\footnote{For
a radiation temperature $T_r\simeq10{\rm KeV}$, the pair
production effect is unimportant (Weaver 1974)}}, as performed by
Weaver (1976), is beyond the scope of the present work. However,
there is a useful numerical criterion for checking the
self-consistency of a radiation-dominated shock. According to
Weaver (1976), the radiative heat transport suffices for producing
the necessary dissipation in a strong shock wave if the ratio of
the radiation pressure to matter pressure $P_r/P_m$ exceeds 4.45.
This condition is satisfied in our case, the ratio of the two
pressures being $P_r/P_m=a T_r^4/n_e kT_e=\rho v_s^2/ n_e k
T_r\simeq 100$ for $v_s=10^9 {\rm cm s^{-1}}$ and $T_e=T_r=10 {\rm
KeV}$, justifying the assumption of a radiation-dominated shock.

The shock-heated walls of the jet channel, before the latter
closes up again, can radiate a transient thermal pulse with
characteristic photon energy $kT\sim {\rm 10KeV}$ and a luminosity
\begin{equation}
\begin{array}{ll}
L_{X}\simeq \sigma T^4 \pi(R_{\star} \theta_f)^2 \\
=2\times10^{48}
\alpha^{3/2}L_{j,49}^{1/2}\left(\frac{v_h}{c}\right)^{1/2}R_{\star,11}\eta_1^{-2}{\rm
erg s^{-1}}
\end{array}
\end{equation}

{ As the cocoon} escapes from the channel, the cocoon pressure
decreases significantly, as does the radiation energy density of
the photon field. Finally, as the pressure wanes, the channel
closes up again under the pressure of the external stellar
material. { The emission from the channel walls vanishes once the
channel has reclosed, so this transient thermal pulse is expected
to last for a time comparable to the jet crossing time given by
Eq.(10). This could appear as a thermal precursor} of $\sim10$
seconds duration, besides the non-thermal emission component from
the cocoon itself.

\section{Possible formation mechanisms of the precursor jets }
\label{sec:jetmech}

How does the weak precursor jet form during the initial core
collapse? There is a long-standing speculation that when rotation
and magnetic fields are taken into account, the core collapse of
massive stars can lead to some form of MHD outflows, although it
is unknown whether they can power supernovae or gamma-ray bursts
(e.g. LeBlance \& Wilson 1970; Wheeler et al. 2000, 2002; Woosley
\& Bloom 2006; Metzger et al. 2007). Following Wheeler et al.
(2000), here we consider that a relatively weak jet of energy a
few times $10^{50}{\rm erg}$ is produced by a rotating
proto-neutron star during the initial core collapse phase. This is
motivated by the consideration that the rotation energy of a
proto-neutron star with a radius of 50 km is suitable for GRB
precursor jets. Immediately after the core collapse, a
proto-neutron star (PNS) forms with a radius of $\sim50 {\rm km}$
and mass $\sim1.4 M_{\sun}$.  For a specific angular momentum of
the progenitor core of $j\simeq10^{16}{\rm cm^{-2} s^{-1}}$, the
angular rotation velocity of the PNS is $\Omega\simeq 400 {\rm
s^{-1}}$, assuming angular momentum conservation. The rotation
energy of the proto-neutron star is
\begin{equation}
E_{\rm rot, PNS}\simeq \frac{1}{2}I_{\rm PNS} \Omega^2\simeq
2\times10^{51}\left(\frac{\Omega}{400 {\rm s^{-1}}}\right)^2
\left(\frac{R_{\rm PNS}}{50{\rm km}}\right)^2 {\rm erg}
\end{equation}
Over a cooling time $\sim 5-10{\rm s}$, the hot proto-neutron star
deleptonizes by neutrino emission and contracts to form the final
neutron star.  If a fraction, e.g. $\sim10\%$ of the rotation
energy is tapped by some MHD process and converted into a
relativistic outflow, this outflow will have the right energy and
timescale for a GRB precursor. One possible scenario for such an
MHD process is that the differentially rotating neutron star can
wind up the poloidal seed field inside the proto-neutron star into
strong toroidal fields, which then emerge from the star through
buoyancy (Klu\'{z}nizk \& Ruderman 1998; Dai \& Lu 1998; Dai et
al. 2006). Once the field emerges through the stellar surface, it
will trigger an explosive reconnection and release the magnetic
energy, resulting in a high-entropy fireball.  Assuming a critical
field value required for buoyancy of $B_f\simeq 10^{15}{\rm G}$
corresponding to the density of the proto-neutron star (Wheeler et
al. 2002), the number of revolutions required is $n_f=B_f/2\pi
B_0=160B_{f,15} B_{0,12}^{-1}$ where $B_0$ is the magnetic field
of the proto-neutron star. The amplification timescale before the
field is expelled by buoyancy is thus $\tau_f\simeq n_f P_{\rm
PNS}\simeq 2.5 B_{f,15} B_{0,12}^{-1}{\rm s}$, where $P_{\rm PNS}$
is the spin period of the proto-neutron star.

Note that the infalling matter from the exterior of the
proto-neutron star is insufficient to suppress this buoyant
magnetic field pressure  in a proto-neutron star. The ram pressure
at $R_{\rm PNS}=50 {\rm km}$ for a mass infall rate $\dot{M}$ is
\begin{equation}
p_{\rm ram}=\frac{\dot{M} v_{ff}}{4\pi R_{\rm PNS}^2}\simeq
5\times10^{26}\dot{M}_{-2}M_0^{1/2}\left(\frac{R_{\rm PNS}}{50{\rm
km}}\right)^{-5/2} {\rm erg s^{-1}}
\end{equation}
(Fryer 1999), which is much smaller than the magnetic field
pressure $p_B=B_f^2/8\pi\simeq 4\times10^{28}B_{f,15}^2{\rm erg
s^{-1}}$, where $v_{ff}=(2GM/R_{\rm PNS})^{1/2}$ is the free-fall
velocity , $\dot{M}$ is the mass infalling rate in units of
$M_{\sun} {\rm s^{-1}}$ and $M_0$ is the proto-neutron star mass.
The infall rate of matter at  early times (the first $10$ s) in
the fallback collapsar is uncertain, but as a rough estimate, the
accretion rate ($\sim0.01 M_{\sun} s^{-1}$) during the fallback
phase (the first 100 s) can be used  (see Fig.5 of MacFadyen et
al. 2001). { However, after about 10 s neutrino cooling will tend
to shrink the proto-neutron star size}, and since the ram pressure
is proportional to $R^{-5/2}$, even if a magnetar with $B\la
10^{15} {\rm G}$ could form after contraction, its activity would
be suppressed, due to its smaller size $R$, by the ram pressure of
the infalling matter after the infall time $\sim 100$ s.


{ Since the activity of a  central neutron star with a magnetic
field $B\la10^{15}{\rm G}$  can be suppressed by the infalling
matter, no significant emission would be seen after the
proto-neutron star phase, until the accretion disk forming at a
later time results in the formation of a newborn black hole. In
the fallback collapsar scenario, the accretion rate depends on the
initial supernova shock energy. For a weak SN explosion with an
energy $\la 0.5\times10^{51}{\rm erg}$, the accretion rate is
about $\sim0.01 M_{\sun} s^{-1}$ (see Fig.5 of MacFadyen et al.
(2001)). With the above accretion rate, the neutron star will
collapse to a black hole on a timescale of the order of $30-100$
seconds. After the black hole forms, gas that continues to fall in
with sufficient angular momentum will settle into a disk, which
can power a main jet through MHD processes or neutrino
annihilation  in the disk-black hole system. This $\sim 100$ s
accumulation timescale of the accretion material may be
responsible for the long time delay of inactivity seen in some
bursts with precursors. }

{ Another possible  mechanism for causing a long-time gap might be
the development of instabilities in the hyperaccretion disk (e.g.
Perna et al. 2006; Janiuk et al. 2007). Janiouk et al. (2007)
found that the opaque inner regions of the disk become unstable to
convective motions. This may give rise to the short-term time
variability seen in the prompt phase, but can not explain the
long-time gap. The long time delay may require, e.g., that the
disk is  gravitationally unstable, fragmenting into rings. The
gravitational instability sets in when the Toomre parameter
(Toomre 1964) $Q=c_s \kappa/\pi G \Sigma<1$, where $c_s$ is the
sound speed, $\kappa$ is the local epicyclic frequency and
$\Sigma$ is the mass surface density. The disk in the case of a
merger of compact objects is found to be gravitationally stable,
since  $Q>1$, according} to Lee \& Ramirez-Ruiz (2002) and Lee et
al. (2005). In order to explain the X-ray flares seen in short
GRBs, Perna et al. (2006) speculated that { the outer part of the
disk may become gravitationally unstable, when the accretion rate
is as high as $10 {\rm M_\odot s^{-1}}$.  However, for fallback
collapsars, the outer radius at which this criteria is satisfied
may lie well outside the the accretion shock radius seen in the
fallback calculations by MacFadyen et al.(2001), which casts into
doubt the viability of this mechanism for the long-time gap.}

The long gap may also be produced by a modulated relativistic wind
without postulating any quiet phase in the central engine
(Ramirez-Ruiz et al. 2001). It was also pointed out that an
approach to test this would be through observations of the prompt
afterglow emission from the reverse shock.

\section{Discussion}
\label{sec:discussion}

We have explored the constraints posed by  GRB precursors with
long time gaps on the jet  dynamics while propagating through the
progenitor stars and jet parameters. We showed that a $\sim100$ s
long gap is hard to interpret in terms of a single jet breaking
out through the stellar envelope, and we suggested a possible two
stage scenario, involving a precursor jet  powered by an initial
proto-neutron star, and a main burst powered by a fallback
collapsar leading to a black hole. In this two stage central
engine scenario, the precursor jet in the first stage must be
weak, so as not to disrupt the star, otherwise there would be no
fallback accretion disk to power the second stage main jet. Since
the observed precursor durations are usually short, of the order
of ten seconds, the jet propagation time through the star poses
severe constraints on the jet physical parameters, such as the jet
luminosity and Lorentz factor. One of our aims has been to give an
analytic calculation of the jet propagation time, in order to
explore the properties of precursor gap time models. Comparison of
our analytical results with numerical hydrodynamical simulations
such as those of Woosley \& Zhang (2007) show  that the analytical
dependence of the propagation time on the jet luminosity is in
qualitative agreement with these simulation results.

We have studied the effects of a precursor jet breaking out from
the progenitor and its related possible observational
manifestations, within the context of our two-stage scenario. The
precursor jet opens a channel through the star as it makes its way
out. The channel will re-close after the precursor jet and its
cocoon have escaped, on a timescale of $\sim 10$ s. This allows a
main jet to form as a result of a collapsar black hole accretion,
if the precursor is weak enough so as not to disrupt the star, the
main jet being collimated by the stellar pressure and propagating
outward in the usual manner. These analytical results are
approximate, and numerical hydrodynamical simulation studies of
such a two-stage jet model would be required to verify its
behavior in more detail.

We have considered some of the possible formation mechanisms for
the precursor jet in the framework of the two-stage fallback
collapsar scenario, involving MHD activity of an initial
proto-neutron star. We stress that a proto-neutron star MHD
outflow is just one possibility, especially considering the many
uncertainties involved both in the proto-neutron star and in the
fallback collapsar scenarios. The possible origin of the $\sim 100
$ s inactivity of the central engine, in this two-stage fallback
collapsar framework, is also tentative.  However, for any possible
two-stage scenario, the above discussion on the jet-star
interaction dynamics remains applicable, and the constraints we
have discussed remain applicable.

{ In summary, we have argued that the long time ($\sim 100$ s) gap
between the precursor and the main burst may  be attributed to the
central engine behavior. One natural scenario resulting in a
$\sim$ 100 s delay is the fallback collapsar model (MacFadyen et
al. 2001). We suggest that the initial precursor jet may arise in
an initial stage of the collapsed core, possibly related to MHD
processes or neutrino emission occurring in a proto-neutron star.
We have calculated the crossing time of the weak precursor jet and
have shown that this initial weak jet can break out through the
progenitor envelope in a time less than ten seconds, which is a
typical precursor duration, provided that it has a relatively high
Lorentz factor $\Gamma\ga10$. The channel opened by the precursor
jet along the rotation axis of the collapsar may produce a
transient weak thermal precursor, before it recloses due to the
re-expanding stellar material which was heated by the cocoon of
the precursor  jet. This provides the conditions required for the
collimation of the main jet, which starts out after a typical
fall-back timescale of $\sim 100$ seconds, producing a black hole
leading to the main GRB.}

\acknowledgments We thank the referee for incisive and helpful
reports which have led to significant improvements of the
manuscript. This work is supported in part by NASA NAG5-13286, NSF
AST 0307376, and  the National Natural Science Foundation of China
under grants 10403002 and 10221001, and the Foundation for the
Authors of National Excellent Doctoral Dissertations of China (for
X.Y.W.).

\begin{figure*}
\plottwo{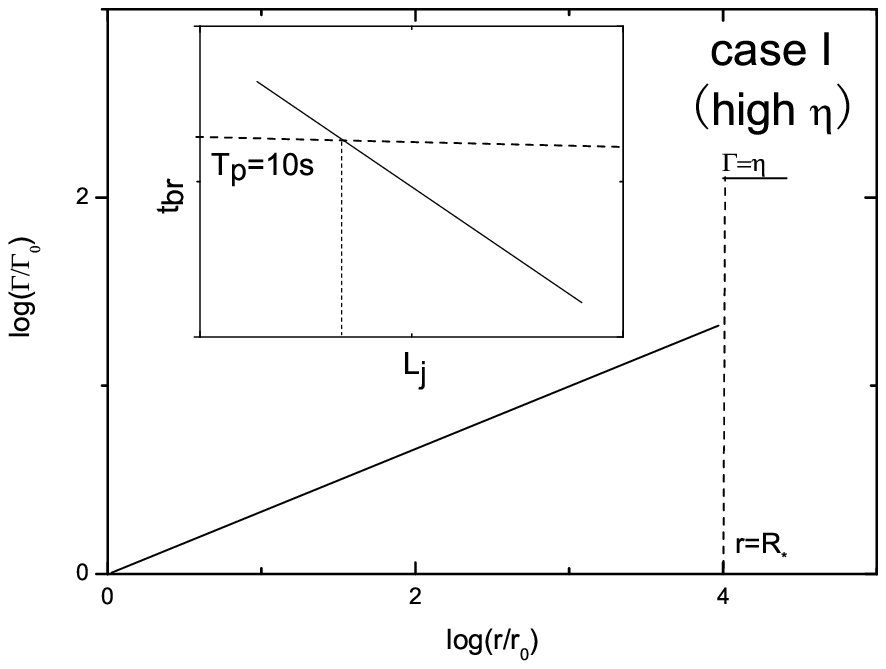} {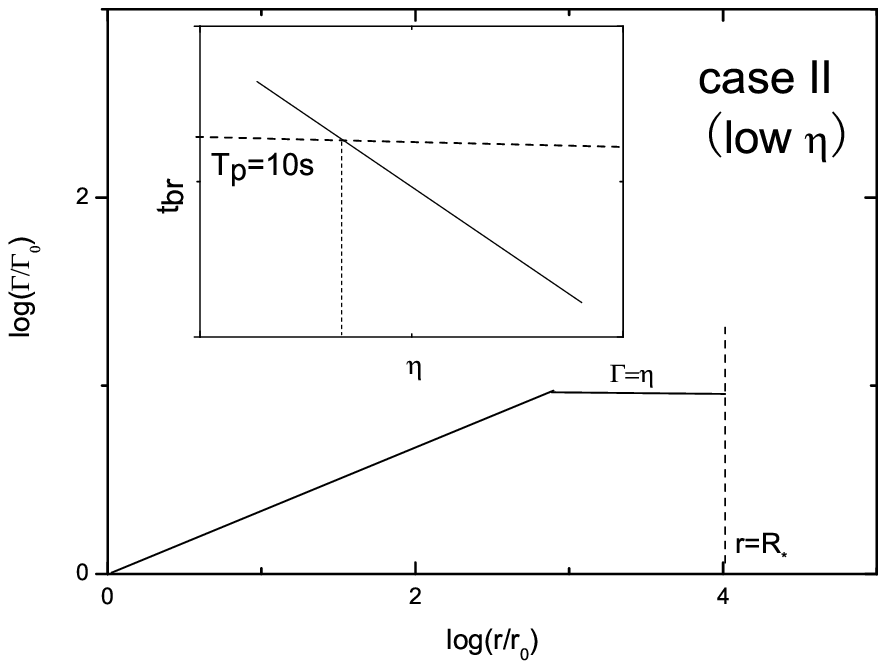} \caption{The evolution of the jet
Lorentz factor with radius for the two different cases defined in
the text. The left panel is the high entropy case, where the jet
Lorentz factor is not saturated before breaking out of the star
and the right one is the lower entropy case, where the jet Lorentz
factor has already saturated before breaking out. The inset
figures in the two panels show the dependence of the jet breakout
time $t_{\rm br}$ on other parameters for the two cases
respectively. Here $r_0$ is the initial injection radius,
$\Gamma_0$ is the initial injection Lorentz factor and $R_*$ is
the stellar radius of the progenitor. }
\end{figure*}

\end{document}